\documentstyle[aps]{revtex}

\def\be{\begin{equation}}
\def\ee{\end{equation}}
\def\bea{\begin{eqnarray}}
\def\eea{\end{eqnarray}}

\makeatletter
\@addtoreset{equation}{section} 
\makeatother

\def\appendix#1
{
\addtocounter{section}{1}\setcounter{equation}{0}
 \renewcommand{\thesection}{\Alph{section}}
 \section*{
 \thesection\protect\indent \parbox[t]{16.715cm}{#1}}
 \addcontentsline{toc}{section}{Appendix \thesection\ \ \ #1}
}

\begin{document}

\title{\begin{flushright}
{\it gr-qc/0405084} \\
\end{flushright}
{\Large {\bf Severe constraints on Loop-Quantum-Gravity
energy-momentum dispersion relation from  black-hole area-entropy law}}}


\author{$~$\\
{\bf Giovanni~AMELINO-CAMELIA}$^b$,
{\bf Michele ARZANO}$^b$
and {\bf Andrea PROCACCINI}$^a$}
\address{$^a$Dipartimento di Fisica, Universit\`{a} di Roma ``La Sapienza''
and INFN Sez.~Roma1,\\
    P.le Moro 2, 00185 Roma, Italy\\
$^b$Institute of Field Physics, Department of Physics and Astronomy,\\
    University of North Carolina, Chapel Hill, NC 27599-3255, USA}

\maketitle

\begin{abstract}\noindent
We explore a possible connection between two aspects of
Loop Quantum Gravity which have been extensively studied in
the recent literature: the black-hole area-entropy law
and the energy-momentum dispersion relation.
We observe that the original Bekenstein argument
for the area-entropy law implicitly requires information
on the  energy-momentum dispersion relation.
Recent results show that in first approximation
black-hole entropy
in Loop Quantum Gravity depends linearly on the area,
with small correction terms which have logarithmic or inverse-power
dependence on the area.
Preliminary studies of the  Loop-Quantum-Gravity dispersion
relation reported some evidence of the presence of terms
that depend linearly on the Planck length, but we observe that
this possibility is excluded since it would require, for consistency,
a contribution to black-hole entropy going like the square root of
the area.
\end{abstract}



\bigskip
\bigskip



\date{\today}


\section{Introduction}
The intuition that
the entropy of a black hole should be proportional
to its (horizon-surface) area, up to
corrections that can be neglected when the area $A$ is much larger
than the square of the Planck length $L_p$,
has provided an important element of guidance for
quantum-gravity research.
It is noteworthy that, as shown by Bekenstein~\cite{bek},
this contribution to black hole entropy
can be obtained from very simple ingredients.
One starts from the general-relativity result~\cite{christo}
that the minimum increase of area when the black hole absorbs a classical
particle of energy $E$ and size $s$ is $\Delta A \simeq 8 \pi L_p^2 E s$
(in ``natural units" with $\hbar=c=1$).
Taking into account the quantum properties of particles
one can estimate $s$ as roughly given by
the position uncertainty $\delta x$, and, since
a particle with position uncertainty  $\delta x$
should at least~\cite{landau} have energy $E \sim  1/ \delta x$,
this leads to the conclusion\cite{bek,hod} that the minimum change
in the black-hole area must be of order $L_p^2$,
independently of the  size of the area.
Then using the fact
that, also independently of the size of the area,
 this minimum increase of area should correspond to the
minimum (``one bit") change of entropy
one easily obtains~\cite{bek} the proportionality between
black-hole entropy and area.

It is remarkable that, in spite of the humble ingredients
of this Bekenstein analysis,
the entropy-area relation
introduced such a valuable constraint for quantum-gravity research.
And a rather challenging constraint, since attempts
to reproduce the entropy-area-linearity result
using directly some quantum properties of black holes
were unsuccessful for nearly three decades.
But over the last few years
both in String Theory and in Loop Quantum Gravity
the needed techniques for the analysis of entropy on the basis of
quantum properties of black holes were developed.
These results~\cite{stringbek,lqgbek1,lqgbek2,lqgbek3} now go even beyond the
entropy-area-proportionality contribution:
they establish that the leading correction should
be of log-area type, so that one expects (for $A \gg L_p^2$)
an entropy-area relation for black holes of the type
\begin{equation}
S = \frac{A}{4 L_p^2} + \rho \ln \frac{A}{L_p^2} + O\left(\frac{L_p^2}{A}\right)
~.
\label{linPLUSlog}
\end{equation}
For the case of Loop Quantum Gravity, which is here of interest,
there is still no consensus on
the coefficient of the logarithmic correction, $\rho$,
but it is established~\cite{lqgbek1,lqgbek2,lqgbek3}
that there are no correction terms
with stronger-than-logarithimic dependence on the area.

We observe that the availability of results on
the log-area correction
might provide motivation for reversing the
Bekenstein argument: the knowledge of black-hole entropy
up to the leading log correction can be used to establish the
Planck-scale modifications of the ingredients of the Bekenstein
analysis.

In particular, the mentioned role of the
relation $E \ge 1/ \delta x$ in the Bekenstein analysis
appears to provide an opportunity to
put under scrutiny some scenarios for the
energy-momentum dispersion relation in Loop Quantum Gravity.
Several recent studies
have tentatively argued that the Loop-Quantum-Gravity
dispersion relation
might involve a term with a linear dependence on the Planck length,
and, as we observe in Section~II,
this in turn requires a Planck-length modification of the
relation $E \ge 1/ \delta x$ between the energy and position uncertainty
of a particle.
However, as we show in Section III, the resulting modification
of the $E \ge 1/ \delta x$ relation would in turn lead,
following the Bekenstein argument, to a contribution to
black-hole entropy that goes like the square root of the area.
Since such a square-root contribution is, as mentioned, excluded
by direct analysis of  black-hole entropy in Loop Quantum Gravity,
we conclude that the presence
in the energy-momentum dispersion relation
of a term with linear dependence on the Planck length is also excluded.

\section{Loop-Quantum-Gravity dispersion relation
and its implications for the $E\ge 1/ \delta x$ relation}
The possibility of Planck-scale modifications of the dispersion
relation has been considered extensively in the recent
quantum-gravity literature~\cite{grbgac,garay,dsr}
and in particular
in Loop Quantum Gravity~\cite{gampul,mexweave,leekoda,kodadsr}.

Some calculations in Loop Quantum Gravity~\cite{gampul,mexweave}
provide support for the idea of
an energy-momentum dispersion relation
that for a particle of high energy
would take the approximate form
\begin{equation}
E \simeq p + \frac{m^2}{2p} + \alpha L_p E^2
~,
\label{disprelONE}
\end{equation}
where $\alpha$ is a coefficient of order 1.
However, these results must be viewed as preliminary~\cite{leekoda,kodadsr}
since they essentially consider
perturbations of ``weave states"~\cite{gampul,mexweave},
rather than perturbations of the ground state of the theory.
It is not surprising (and therefore not necessarily insightful)
that there would be some states of the theory whose
excitations have a modified spectrum.
If instead a relation of the type (\ref{disprelONE})
was applicable to excitations of the ground state of the
theory this would provide a striking
characteristic of the Loop-Quantum-Gravity approach.

Several papers have been devoted  to the derivation of tighter an tighter
experimental limits on coefficients of the $\alpha$ type
for Loop Quantum Gravity (see, {\it e.g.}, Ref.~\cite{sudaPRL}
and references therein). As announced we intend to show here that
the linear-in-$L_p$ term can be excluded already on theoretical grounds,
because of an inconsistency with the black-hole-entropy results.

In this section we start by observing that a modified dispersion
relation implies a modification of the relation $E \ge 1/ \delta x$
between the energy of a particle and its position uncertainty.
We can see this by simply following
the familiar derivation~\cite{landau} of the
relation $E \ge 1/ \delta x$, substituting, where applicable, the
standard special-relativistic dispersion relation with
the Planck-scale modified dispersion relation.
It is convenient to focus first~\cite{landau} on the case
of a particle of mass $M$ at rest, whose position is being measured
by a procedure involving a collision with a photon of energy $E_\gamma$
and momentum $p_\gamma$.
In order to measure the particle position with precision $\delta x$
one should use a photon with
momentum uncertainty $\delta p_\gamma \ge 1/\delta x$.
Following the standard argument~\cite{landau},
one takes this $\delta p_\gamma \ge 1/\delta x$ relation and converts it into
the relation $\delta E_\gamma \ge 1/\delta x$, using the
special-relativistic dispersion relation,
and then the relation $\delta E_\gamma \ge 1/\delta x$ is converted into
the relation $M\ge 1/\delta x$ because the measurement procedure
requires\footnote{One must take into account the fact~\cite{landau}
that the measurement procedure should ensure
that the relevant energy uncertainties are not large enough
to possibly produce extra copies of the particle whose
position one intends to measure.} $M \ge \delta E_\gamma$.
If indeed Loop Quantum Gravity hosts a Planck-scale-modified
dispersion relation of the form (\ref{disprelONE}),
it is easy to see that, following the same reasoning,
one would obtain from $\delta p_\gamma \ge 1/\delta x$
the requirement $M  \ge (1/ \delta x) [1 + 2 \alpha (L_p / \delta x)]$.

These results strictly apply only to the measurement of the position of
a particle at rest, but they can be straightforwardly
generalized~\cite{landau} (simply using a boost)
to the case of measurement of the position of a particle of energy $E$.
In the case of the standard dispersion relation (without Planck-scale modification)
one obtains the familiar $E  \ge 1/\delta x$.
In the case of (\ref{disprelONE}) one instead easily finds that
\begin{equation}
E  \ge \frac{1}{\delta x} \left(1 + 2 \alpha \frac{L_p}{\delta x}\right)
~.
\label{deform}
\end{equation}

\section{A requirement of consistency with the black-hole entropy
analysis}
We now intend to show that the linear-in-$L_p$ modification
of the relation
between the energy of a particle and its position uncertainty,
which follows from the corresponding modification of the
energy-momentum dispersion relation,
should be disallowed in Loop Quantum Gravity since it leads to
a contribution to the black-hole entropy-area relation which
has already been excluded in direct black-hole-entropy analyses.

We do this by following
the original Bekenstein argument~\cite{bek}.
As done in Ref.~\cite{bek}
we take as starting point the general-relativistic
result which establishes that the area
of a black hole changes according to $\Delta A \ge 8 \pi E s$
when a classical particle of energy $E$ and size $s$ is absorbed.
In order to describe the absorption of a quantum particle one must
describe the size of the particle in terms of the uncertainty
in its position~\cite{bek,hod}, $s \sim \delta x$, and take into account
a ``calibration\footnote{Clearly some calibration
is needed in order to adapt
the classical-gravity result for absorption of a classical particle
to the case of a quantum black hole absorbing a quantum particle.
In particular, a calibration should arise
in the description of a quantum particle with position
uncertainty $\delta x$ in terms of a classical particle of size $s$.
A direct evaluation of the calibration coefficient
within quantum gravity is presently beyond reach;
however, several authors (see, {\it e.g.}, Refs.~\cite{chenproc,calib2,calib3})
have used the independent analysis of black-hole
entropy by Hawking~\cite{hawkBH} to infer indirectly this calibration
needed in the Bekenstein argument.
We adopt this calibration for consistency with previous literature,
but the careful reader will notice that this calibration
does not affect our line of analysis (the calibration could be reabsorbed in
the free parameter $\alpha$).}
factor"~\cite{chenproc,calib2,calib3} $(\ln 2)/2 \pi$
that connects the $\Delta A \ge 8 \pi E s$ classical-particle result
with the quantum-particle
estimate $\Delta A \ge 4 (\ln 2) L_p^2 E \delta x$.
Following the  original Bekenstein argument~\cite{bek} one then
enforces the relation $E \ge 1/ \delta x$
(and this leads to $\Delta A \ge 4 (\ln 2) L_p^2$),
but we must take into account the Planck-length modification in (\ref{deform}),
obtaining
\begin{eqnarray}
\Delta A \ge 4 (\ln 2) \! \left[L_p^2  + \!  2 \frac{\alpha L_p^3}{\delta x}
 \right] \! \simeq
 4 (\ln 2) \! \left[ L_p^2  + \! 2 \frac{\alpha L_p^3}{R_S}
  \right]
\! \simeq  4 (\ln 2) \! \left[ L_p^2
+ \! \frac{\alpha 4 \sqrt{\pi} L_p^3}{\sqrt{A}}
 \right] ~,
\nonumber
\end{eqnarray}
where we also used the fact that in falling in the black hole
the particle acquires~\cite{calib2,smaller,bigger} position
uncertainty $\delta x \sim R_S$, where $R_S$ is the Schwarzschild
radius (and of course $A = 4 \pi R_S^2$).

Next, following again Bekenstein\cite{bek}, one assumes that the entropy
depends only on the area of the black hole, and one uses the fact
that according to information theory
the minimum increase of entropy should be $\ln 2$,
independently of the value of the area:
\begin{equation}
\frac{dS}{dA} \simeq \frac{min (\Delta S)}{min (\Delta A)}
\simeq  \frac{\ln 2}{4 (\ln 2)  L_p^2 \left[1
+ \alpha 4 \sqrt{\pi} \frac{L_p}{\sqrt{A}} \right]}
\simeq  \left(\frac{1}{4 L_p^2}
-  \frac{\alpha \sqrt{\pi}}{ L_p \sqrt{A}} \right)
~.
\label{minDa}
\end{equation}
From this one easily obtains (up to an irrelevant constant contribution
to entropy):
\begin{equation}
S \simeq \frac{A}{4 L_p^2}
- 2 \alpha  \sqrt{\pi} \frac{\sqrt{A}}{L_p}
~.
\label{final}
\end{equation}

We therefore conclude that
when a quantum-gravity theory predicts the presence of a linear-in-$L_p$
contribution to the energy-momentum dispersion relation
it should correspondingly predict the presence of $\sqrt{A}$
contribution to  black-hole entropy.
Since in Loop Quantum Gravity such a $\sqrt{A}$
contribution to  black-hole entropy has already been
excluded~\cite{lqgbek1,lqgbek2,lqgbek3}
in direct black-hole entropy studies, we conclude that
in Loop Quantum Gravity
the presence of linear-in-$L_p$
contributions to the energy-momentum dispersion relation
is excluded.

It is instead plausible that Loop Quantum Gravity might host
a dispersion relation of the type
\begin{equation}
E \simeq p + \frac{m^2}{2p} + {\tilde {\alpha}} L_p^2 E^3
~,
\label{disprelTWO}
\end{equation}
with a quadratic-in-$L_p$ contribution.
In fact, the careful reader can easily adapt our analysis
to the case of the
dispersion relation (\ref{disprelTWO}), finding
that the quadratic-in-$L_p$ contribution to the dispersion relation
ultimately leads to a leading correction to the black-hole-entropy formula
which is of log-area type,
consistently with the indications obtained
in direct black-hole entropy studies~\cite{lqgbek1,lqgbek2,lqgbek3}.

\section*{Acknowledgments}
G.~A.-C.~gratefully acknowledges conversations
with O.~Dryer, D.~Oriti, C.~Rovelli and L.~Smolin.
The work of M.~A.~was supported by a Fellowship from The Graduate School of The
University of
North Carolina. M.~A.~also thanks the Department of Physics of the University of
Rome for hospitality.

\baselineskip 12pt plus .5pt minus .5pt

{\small

\end{document}